\documentclass[manuscript]{aastex}

\shorttitle{NH$_3$ absorption in the visible spectrum of Jupiter}
\shortauthors{Irwin et al.}

\begin{document}


\title{Analysis of gaseous ammonia (NH$_3$) absorption in the visible spectrum of Jupiter - Update}


\author{Patrick G. J. Irwin, Neil Bowles, Ashwin S. Braude, Ryan Garland, and Simon Calcutt }
\affil{Department of Physics, University of Oxford, Parks Rd, Oxford OX1 3PU, UK.}
\author{Phillip A. Coles, Sergey N. Yurchenko and Jonathan Tennyson}
\affil{Department of Physics and Astronomy,  University College London, London WC1E 6BT, UK}
\email{patrick.irwin@physics.ox.ac.uk}



\begin{abstract}
An analysis of currently available ammonia (NH$_3$) visible-to-near-infrared gas 
absorption data was recently undertaken by Irwin et al. (Icarus, 302 (2018) 426) to help interpret Very 
Large Telescope (VLT) MUSE observations of Jupiter from 0.48 -- 0.93 $\mu$m, made in 
support of the NASA/Juno mission. Since this analysis a newly revised set of 
ammonia line data, covering the previously poorly constrained range 0.5 -- 0.833 $\mu$m, has been released by the ExoMol project, ``C2018" \citep{jtNH3PES}, which demonstrates 
significant advantages over previously available data sets, and providing for the first time complete line data for the previously poorly constrained  5520- and 6475-\AA\ bands of NH$_3$. In this paper we 
compare spectra calculated using the ExoMol--C2018 data set \citep{jtNH3PES} with spectra calculated 
from previous sources to demonstrate its advantages. We conclude that at the 
present time the ExoMol--C2018 dataset provides the most reliable ammonia absorption source 
for analysing low- to medium-resolution spectra of Jupiter in the visible/near-IR spectral range, but note that the data are less 
able to model high-resolution spectra owing to small, but significant 
inaccuracies in the line wavenumber estimates. This work is of significance not 
only for solar system planetary physics, but for future proposed observations of 
Jupiter-like planets orbiting other stars, such as with NASA's planned 
Wide-Field Infrared Survey Telescope (WFIRST). 
\end{abstract}





\section{Introduction}

In a recent paper \citep{irwin18} we reported an analysis of the currently 
available sources of gaseous ammonia (NH$_3$) absorption data to model 
observations of Jupiter we have been making with the MUSE (Multi Unit 
Spectroscopic Explorer, \cite{bacon10}) instrument at ESO's (European Southern 
Observatory) Very Large Telescope (VLT), in support of the NASA/Juno mission. We 
found that the ammonia k-tables generated from the band models of 
\cite{bowles08} provided the best combination of reliability and wavelength 
coverage for the MUSE spectral range, although these data were found to become very noisy at wavelengths less than 0.758 $\mu$m, leading to uncertain absorption coefficients, and did not cover the bands at 0.648 and 0.552 $\mu$m. We also found that the data of 
\cite{bowles08} seemed consistent with the ExoMol -- BYTe ammonia line data of 
\cite{yurchenko11}, where they overlap (0.8 - 1.05 $\mu$m), but that the 
existing BYTe data did not cover the ammonia absorption bands at 0.79 and 
0.765 $\mu$m, which are prominent in our MUSE observations. At shorter 
wavelengths we found that the laboratory observations of \cite{lutzowen80} 
provided a good indication of the position and shape of the ammonia absorptions 
near 0.552 $\mu$m and 0.648 $\mu$m, but their absorption strengths seemed 
inconsistent with the available data at longer wavelengths and there was no 
reliable way to extrapolate the strength and shape of these bands to the cold, 
H$_2$-He broadening conditions in Jupiter's atmosphere. Finally, we concluded 
that the line data of the 0.648-$\mu$m band of \cite{giver75} were not suitable 
for modelling these data as they accounted for only 17\% of the band absorption 
and lacked information on lower state energies, necessary to compute their 
absorption strengths at low temperatures. 

In this paper we compare these data sources with a newly released room temperature ammonia line list, ``C2018" \citep{jtNH3PES} from the ExoMol  project \citep{jt528,jt631}. The new line list covers an extended wavelength range compared to BYTe \citep{yurchenko11}, such that the shortest wavelength covered has been reduced from 0.833 to 0.5 $\mu$m 
(i.e. upper wavenumber increased from 12000 cm$^{-1}$ to 20000 cm$^{-1}$) and 
find that the revised ExoMol line data,  C2018, provide a very good and reliable resource 
for modelling our VLT/MUSE observations and indeed all future visible/near-IR observations of Jupiter.

\section{New ExoMol ammonia line data}

The previously existing ExoMol \citep{jt528,jt631} ammonia line data set, BYTe
\citep{yurchenko11}, contained over 1 billion lines based on an empirically-tuned
potential energy surface (PES), and an \textit{ab initio} dipole moment surface and variational
nuclear motion calculations. Tuning of the PES involved adding a perturbation term to the \textit{ab initio} surface, which is then varied through a least-squares-fitting procedure to both experimentally derived energies and \textit{ab initio} electronic energies \citep{jt503}. By this process the accuracy of the PES is drastically improved, whilst still retaining a form that closely resembles the \textit{ab initio} surface.

BYTe covers the spectral range 0 -- 
12000 cm$^{-1}$ (i.e. wavelengths longer than 0.833 $\mu$m), and includes line 
strengths and lower state energies, but does not have any information on line 
widths. Hence, for our previous study of available ammonia absorption data \citep{irwin18}, we had to undertake further analysis and assumptions and assign 
line widths (under H$_2$/He-broadening conditions) and their temperature 
dependences in order to apply these data to radiative transfer calculations 
under Jovian conditions. Most of the lines are `hot lines' that only become 
important at high temperature (i.e. $T > 400$ K) and are not important at Jovian 
temperatures. Hence, we initially reduced the number of lines to a more 
manageable level by using the lower state energies and partition function to 
compute the line strengths at 400 K and neglected lines contributing less than 
$10^{-4}$ \% to the total line strength\footnote{summed over 1 cm$^{-1}$-wide bins}.  
Once we had reduced the number of lines, we then needed to add line-broadening 
information since (unless the lines are to be used for calculations at low 
pressure and high temperature conditions, where only Doppler-broadening of the lines is 
important) we need information on the pressure-broadened line widths to compute 
the Voigt lineshape for each line. Following \cite{amundsen14} and 
\cite{garland17}, the foreign-broadening line widths of the NH$_3$ lines in a 
solar composition H$_2$-He (assumed 85:15 ratio) atmosphere were allocated 
from the data of \cite{pine93} (depending on the rotational energy level J), 
while the temperature-dependence exponents of these widths were taken from 
\cite{nouri04} and \cite {sharp07} for H$_2$-broadening and He-broadening, 
respectively,  and used to compute a weighted-average of 0.66 \citep{garland17}. 
For self-broadening, the line widths were taken from \cite{markov93} (not \cite{pine93} as was mistakenly reported by \cite{irwin18}), while 
the temperature dependence exponent was set to the default expected theoretical 
value of 0.5\footnote{The collisioned-broadened width, $\gamma$, is proportional to 1/(time between collisions), 
or equivalently the thermal velocity divided by the mean-free-path. Since thermal velocity, $v \propto \sqrt{T}$ and 
mean free path, $\lambda \propto$ 1/density, or equivalently $\lambda \propto T/p$, we find $\gamma \propto p/\sqrt{T}$, giving a temperature dependence exponent of 0.5.}. 

The newly computed room-temperature ExoMol ammonia line list, C2018 \citep{jtNH3PES}, covers the 0 -- 
20000 cm$^{-1}$ range (i.e. wavelengths longer than 0.5 $\mu$m), and, for this work,  was provided in a HITRAN-like format with line widths 
and temperature dependences included from a number of different sources, which are reviewed by \cite{wilzewski15}.
The H$_2$-broadening widths were taken from the polynomial fit by \cite{nemtchinov04} up to $J = 9$, and a value of 0.0788 cm$^{-1}$ atm$^{-1}$ assumed beyond this, as used by \cite{wilzewski15}. The He-broadening widths were taken from \cite{wilzewski15}, who used the unpublished polynomial fit by Linda R. Brown up to $J = 9$ and a value of 0.0282 cm$^{-1}$ atm$^{-1}$ for higher $J$ values. The self-broadening coefficients were also taken from a polynomial fit by \cite{nemtchinov04} up to $J = 8$, and a value of 0.5 cm$^{-1}$ atm$^{-1}$ assumed for higher J-values, as given in HITRAN2016 \citep{gordon17}. The temperature dependence of the H$_2$-broadening was taken from \cite{wilzewski15}, who used the polynomial fit by \cite{nemtchinov04} up to $J = 9$ and a value of 0.59 thereafter. The temperature dependence of the He-broadening was estimated to be 0.37 as given by \cite{wilzewski15},  who averaged the data from several sources. The temperature dependence of self-broadening was taken to be  0.79 by averaging the measured values of \cite{baldacchini00}.

Compared to BYTe, the C2018 line list \citep{jtNH3PES} is based on an improved, empirical potential energy surface 
and a significantly larger ro-vibrational basis set used in the diagonalization of the molecular Hamiltonian. These improvements are
necessary for the extension to short wavelengths. Using the same spectroscopic model, 
a hot line list complete up to 1500 K for frequencies below 12000 cm$^{-1}$, and up to 300 K for the 12000$-$20000 cm$^{-1}$ range \citep{jtCoYuTe}, 
is under construction and will eventually contain several billion lines. This will be the full ``CoYuTe" line list and, as well as being suitable for higher temperatures, will have the calculated energies replaced by their corresponding experimental values if and where possible. These energies are all below 7555 cm$^{-1}$, and for these lines there will be a very small improvement on some line positions ($ < 0.1$ cm$^{-1}$). However, these improvements will not affect calculations in the visible/near-IR range at temperatures less than 400 K and so in the range we are considering in this paper the C2018 and CoYuTe data sets are effectively identical.

\begin{figure}
\epsscale{0.6}
\plotone{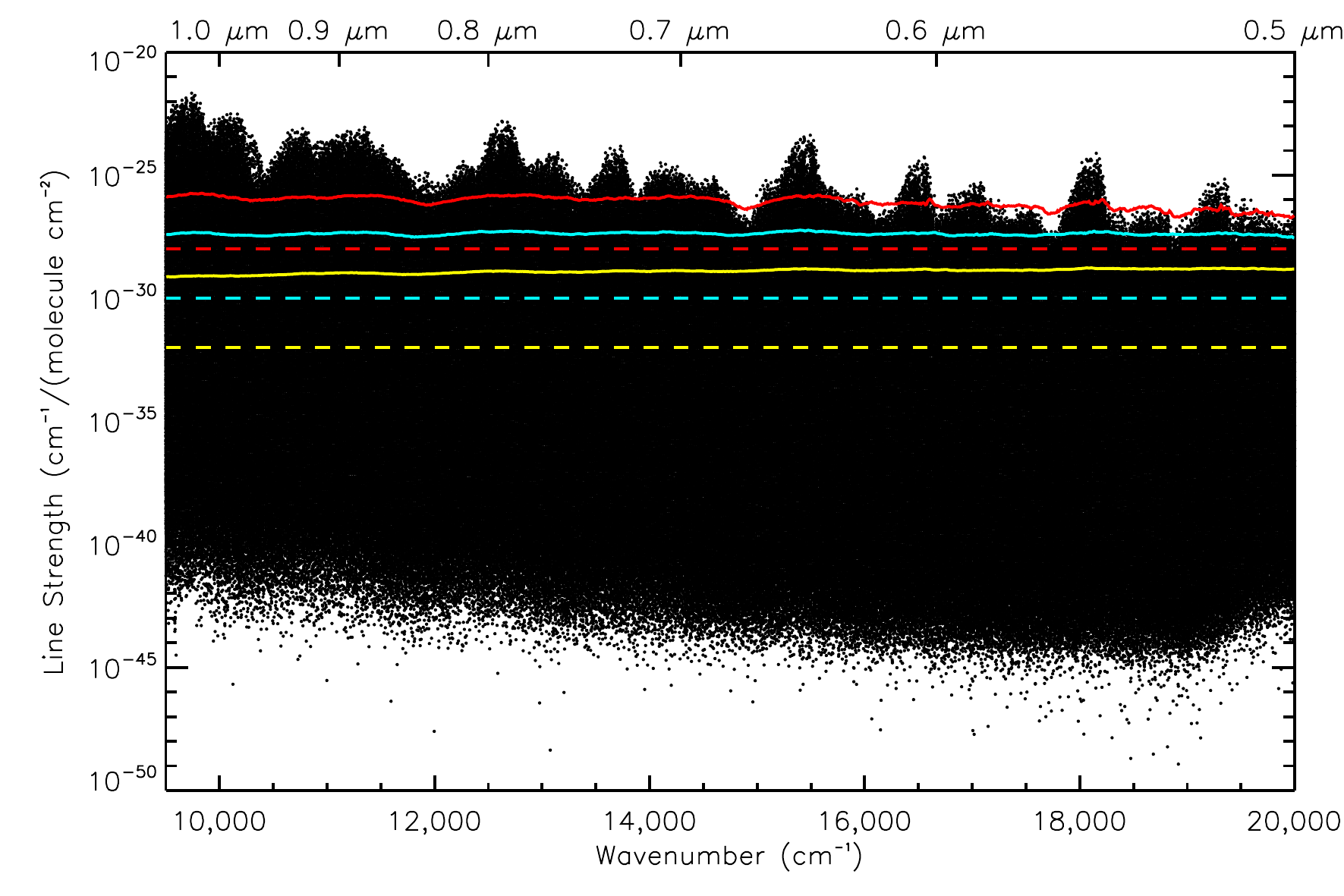}
\caption{Summary of lines in the ExoMol--C2018 line database \citep{jtNH3PES}. The strengths of all lines (at 296 K) are shown between 9500 and 20000 cm$^{-1}$. Equivalent wavelengths are indicated on the top axis. The horizontal, dashed lines indicate the different line strength cut-offs that were explored of $1 \times 10^{-28}$,  $1 \times 10^{-30}$ and  $1 \times 10^{-32}$ cm$^{-1}$/(molecule cm$^{-2})$, respectively. The corresponding solid lines show the integrated `pseudo continuum' absorption of all the omitted lines for the different cut-off strengths. \label{CoYuTesummary}}
\end{figure}

\begin{figure}
\epsscale{0.8}
\plotone{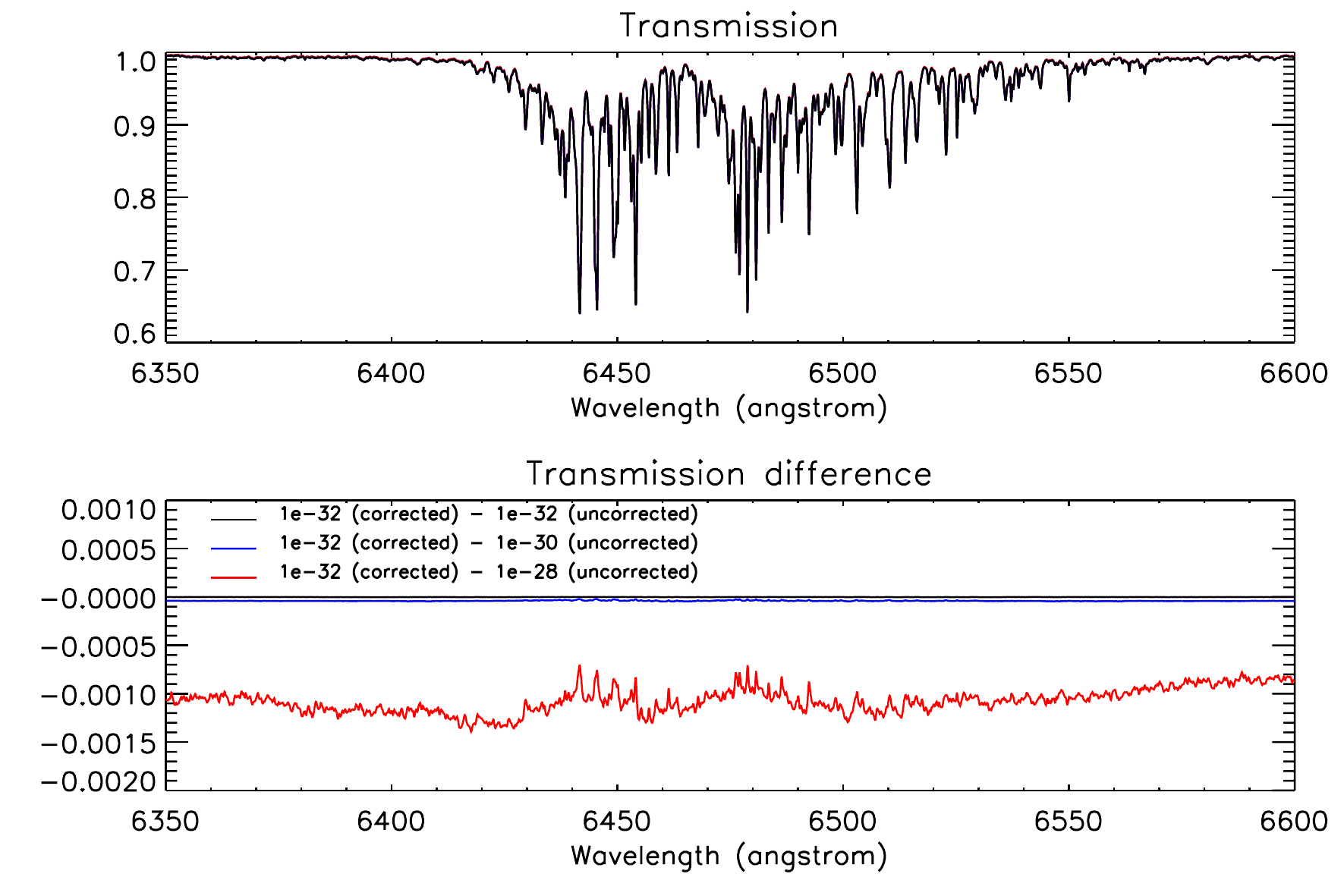}
\caption{Line-by-line-calculated transmission spectrum of the self-broadened path of ammonia shown in Fig. 3 of \cite{giver75}, with spectral resolution of 0.2\AA, using the new ExoMol--C2018 linedata \citep{jtNH3PES}. In this calculation the temperature is 294 K, the path length is 36 m and the pressure is 1 atm. The top plot compares the calculated spectra using a line strength cut-off of $1 \times 10^{-32}$ cm$^{-1}$/(molecule cm$^{-2})$ including `pseudo line-continuum' correction (black) with spectra calculated with no line continuum correction and higher line strength cutoffs of $1 \times 10^{-30}$ (blue) and $1 \times 10^{-28}$ (red), respectively. As can be seen the transmission calculated with a cut-off of $1 \times 10^{-28}$ and no continuum correction can just be differentiated from the other calculated spectra. The bottom plot compares the difference spectra calculated with different line strength cut-offs, including or excluding the continuum correction. Significant differences are seen for a cut-off of $1 \times 10^{-28}$ cm$^{-1}$/(molecule cm$^{-2})$, but very small difference is seen for $1 \times 10^{-30}$and effectively no difference seen for $1 \times 10^{-32}$, when the continuum correction is omitted. \label{test6475}}
\end{figure}

Although the room-temperature line list used here, C2018, contains far fewer lines than the final high-temperature CoYuTe line database \citep{jtCoYuTe}, it still comprises over 290 million lines in the 9500 - 20000 cm$^{-1}$ range. The line strengths listed in C2018 in this spectral range are summarised in Fig. \ref{CoYuTesummary}. This is far too large a number of lines to realistically use with current analysis techniques. As a first approximation, it is possible to eliminate all lines with a strength of less than a certain cut-off, as we did previously with the ExoMol--BYTe set \citep{irwin18}, and initially all lines with a strength (at 296K) of less than $1\times 10^{-28}$ cm$^{-1}$/(molecule cm$^{-2})$ were neglected, which reduced the number of lines in the 9500 - 20000 cm$^{-1}$ spectral range to 1.3 million. Our initial test calculations were made with this table, but we were concerned whether it was reliable to simply ignore so many lines. Hence we generated two further subsets, eliminating all lines with a strength less than $1\times 10^{-30}$  and $1\times 10^{-32}$ cm$^{-1}$/(molecule cm$^{-2})$,  respectively, which contained 8.8 million and 38 million lines each. Calculating test transmission spectra in the 6475 \AA\ band (Fig. \ref{test6475}), we found that there were very small differences between spectra calculated with a lower strength cut-off of $1\times 10^{-28}$ and $1\times 10^{-30}$, respectively, but negligible difference going from $1\times 10^{-30}$ to $1\times 10^{-32}$. However, as the mean strength of the lines in C2018 reduces as we go to higher wavenumbers we were concerned that a cut-off of even $1\times 10^{-30}$ might not be suitable at shorter wavelengths. In addition, we found that the size of the database for a strength cut-off of $1\times 10^{-30}$ (nearly 9 million lines) led to excessively long computation times when converting these data to the k-tables necessary to model Jovian reflectance spectra. Hence, we looked to see whether it might be possible to more generally, reliably and efficiently model the small, but potentially significant effect of the numerous weak lines in databases such as C2018 and CoYuTe. 

To do this we first of all analysed the full  C2018 line database in wavenumber bins of a certain 
`medium' size, which we chose to be $\Delta\tilde{\nu} = 10$ cm$^{-1}$. The 
lines in each bin were then read in and, if their strength exceeded a chosen minimum, were copied to a new database as before. 
However, the remaining lines, which would otherwise have been rejected, were instead combined into a set of `pseudo line-continuum' 
parameters, computed for each bin: A) $S_T(T_C) = \sum{S_i(T_C)}$, the 
sum of the line strengths for all of the omitted lines: B) $\bar{E}_l$, the 
line-strength-weighted mean lower state energy of the omitted lines (i.e. 
$\bar{E}_l = \sum{S_i(T_C) E_i}/\sum{S_i(T_C)}$); C) $\bar{\gamma}_f$, the 
line-strength-weighted mean foreign-broadened line width of these lines; D) 
$\bar{\gamma}_s$, line-strength-weighted mean self-broadened line width of these 
lines; E) $\bar{n}_f$, the line-strength-weighted mean temperature dependence 
of the foreign-broadening coefficients of these lines; and F) $\bar{n}_s$, the line-strength-weighted mean temperature dependence 
of the self-broadening coefficients of these lines.

When calculating the subsequent contribution of these weak
lines to the continuum at wavenumber, $\tilde{\nu}$, and calculation temperature, $T$, we first calculated the cumulative strength of the weak lines in each bin as:
$$S_T(T)=S_T(T_C) {{Q_V(T_C)Q_R(T_C)(1-\exp(-E_T/kT))\exp(-\bar{E}_l/kT)}\over
{Q_V(T)Q_R(T)(1-\exp(-E_T/kT_C))\exp(-\bar{E}_l/kT_C)}}\eqno(1)$$
where $Q_V(T)$ and $Q_R(T)$ are the vibration and rotation partition functions, 
respectively, of the gas in question, $(1-\exp(-E_T/kT))$ is a correction for 
stimulated emission, equal to one minus the ratio of Boltzmann populations for 
the two states ($E_T$ is the transition energy, equal to $100hc\tilde{\nu}_0$, 
where $\tilde{\nu}_0$ is the wavenumber at the centre of the bin), and 
$\exp(-\bar{E}_l/kT)$ is the Boltzmann population of line-strength-weighted mean 
energy $\bar{E}_l$.  We decided to 
model the effect of our pseudo line-continuum parameters not just within the 
individual bins themselves, but also in adjacent bins. This can be important 
when we have neighbouring bins containing weak lines of very different cumulative strength under conditions of 
considerable line broadening, where we might see the wings of the stronger lines from one bin 
noticeably spilling into adjacent bins with significantly lower cumulative strength. Assuming the weak lines to be equally distributed throughout the bin, $n$, 
with central wavenumber $\tilde{\nu}_n$, we calculate the contribution of 
these absorptions in a neighbouring bin, $i$, with central  wavenumber, 
$\tilde{\nu}_i$, to be
$$C_{ni} (T)= {{S_n(T) V(x_{ni},y)}\over{\sum_i{V(x_{ni},y)}}} \eqno(2)$$
where $V(x_{ni},y)$ is the line-shape function (usually Voigt, including or 
excluding line wing corrections), $x_{ni} = \tilde{\nu}_i-\tilde{\nu}_n$,  and 
$y = \gamma_L/\gamma_D$, where, assuming foreign-broadened conditions, the pressure-broadening line width is
$$\gamma_L = \bar{\gamma}_f {p\over p_C} {\Bigg ({T_C\over T} \Bigg 
)}^{\bar{n}_f}\eqno(3)$$
and the Doppler-broadened line width is
$$ \gamma_D = {{{\tilde{\nu}}_0} \over c} {\Bigg ({2RT \over M_r} \Bigg 
)}^{1/2}\eqno(4)$$
where $M_r$ is the Molecular Weight of the gas in question. These 
pseudo line-continuum contributions are then added and the mean 
continuum absorption at each bin centre calculated to be:
$$\bar{k}_n(T) = \sum_i{C_{ni}}(T)/\Delta\tilde{\nu}\eqno(5)$$

\begin{figure}
\epsscale{0.6}
\plotone{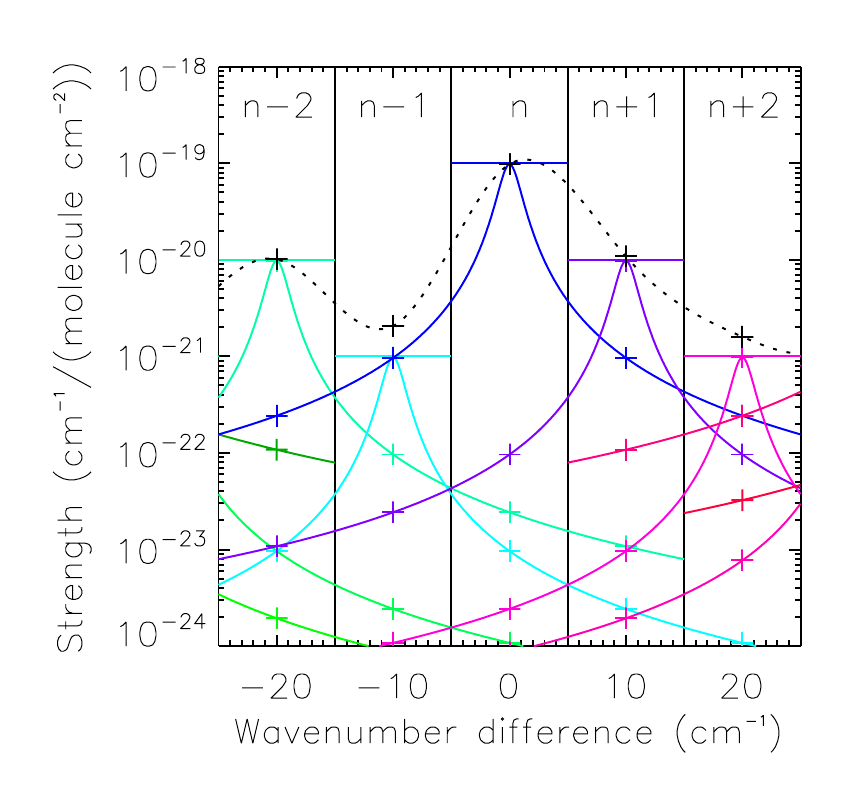}
\caption{Suggested `pseudo line-continuum' correction to deal with very large number of weak lines present in line databases such as ExoMol--CoYuTe \citep{jtCoYuTe}. Here we focus on five adjacent bins of width 10 cm$^{-1}$ and have arbitrarily set widely varying strengths to demonstrate the principle of the suggested scheme. The coloured, horizontal lines indicate the assigned integrated line strengths in each bin. The contribution of these lines to the continuum absorption in adjacent bins is computed using the line shape required and indicated by the coloured solid lines. Here, for simplicity, we have assumed a Lorentz line shape with FWHM = 1 cm$^{-1}$. We have also assumed a line wing cut-off of 35 cm$^{-1}$. For each bin we calculate the contribution to other bins within the line wing cut-off at the bin centres and then normalise to ensure these contributions sum to equal the integrated line strength in the original bin; the value at the bin centres are shown by the coloured crosses. The sum of contributions at each bin centre is then added to give the total in each bin indicated by the black crosses. A cubic spline is then fitted to the integrated log line strengths to allow smooth interpolation across the wavenumber.  When calculating the continuum absorption in a spectrum these adjusted line strengths need to be divided by the width of the bin, here  10 cm$^{-1}$, to turn them into absorption coefficients. \label{figlco}}
\end{figure}

We show this process graphically in Fig. \ref{figlco}, where we focus on five adjacent bins 
of width $\Delta\tilde{\nu} = 10$ cm$^{-1}$; here we have arbitrarily set 
different cumulative weak line strengths in the bins, indicated by the coloured horizontal lines. The contribution to the continuum absorption of the lines in one 
bin to the same bin and adjacent bins (within any line wing cutoff) is computed 
using the required line shape and the calculated values at the  bin centres 
normalised to ensure they sum to the integrated line strength in the original 
bin. The total contribution at the bin centres from the bin itself and adjacent 
bins is then added and these values interpolated to determine the final 
continuum spectrum. We find that using these equations to calculate the contribution of the 
lines that were previously simply discarded gives excellent correspondence with calculations that omit no 
lines at temperatures close to $T_C$, but discrepancies can arise as 
$ | T-T_C |$ becomes greater. Hence, since our k-tables are calculated over a wide range of 
temperatures, we computed pseudo line-continuum parameters at each k-table temperature separately, thus avoiding entirely this temperature extrapolation error. 

Finally, we also explored using the same system to assign to the `pseudo line continuum' lines that were not amongst the top N strongest lines in the bin, or lines contributing less than a certain minimum percentage of the total integrated line strength. This may prove eventually to be more efficient and adaptable than simply assigning all lines with a strength less than a certain cut-off line strength, as we have done here, but development issues led us to use the simpler minimum strength cut-off system described here. We may return to a more general approach in future work.
 
Taking the C2018 line database, we computed a set of reduced line databases and pseudo line-continuum tables (in 10 cm$^{-1}$-wide bins) at a range of temperatures equally spaced between 50 and 400K with a cut-off line strength at each calculation temperature of $1\times 10^{-28}$ cm$^{-1}$/(molecule cm$^{-2})$. These line databases and continuum tables were then used to 
compute k-distribution look-up tables to model our VLT/MUSE observations for H$_2$-He-broadening conditions (assuming H$_2$:He = 0.865:0.135), covering the spectral range 0.4 
-- 1.0 $\mu$m. These k-tables are used in our correlated-k radiative transfer 
model, NEMESIS, described later. In these k-tables, 10 g-ordinates were used 
with 15 pressures logarithmically spaced between 10$^{-4}$ and 10 bar and the 20 
temperatures, mentioned before, equally spaced between 50 and 400K. 


\section{Comparison with Lutz and Owen (1980) and Giver (1975) absorption data}

\cite{lutzowen80} report room-temperature laboratory measurements of the 
absorption spectrum of NH$_3$ for both the 6475 \AA\ band and the 5520 \AA\ band 
at a quoted spectral resolution of 2~\AA. These data are presented in the form of 
apparent absorption cross-sections. We found in our previous analysis 
\citep{irwin18} that these data provided a good estimate of the shape of the 
6475 \AA\ and 5520 \AA\ bands in the observed VLT/MUSE observations, but that 
the estimated strengths under Jovian conditions were not consistent with the 
strength of ammonia gas absorption features calculated at longer wavelengths 
with line data. We attributed this discrepancy to the fact that \cite{lutzowen80} 
 report observations under self-broadened conditions, rather than 
H$_2$--He-broadened conditions, and only for a single temperature with no simple means to extrapolate to lower temperatures. 

In addition, we analysed line data extracted from low- and high-resolution (R= 
$\lambda/\Delta \lambda = 170,000$) laboratory spectra of self-broadened ammonia 
in the 6475 \AA\ band at room temperature by \cite{giver75}. In this line 
database the strengths of the thirty three strongest lines in the 6475 \AA\ band 
were estimated as were their self-broadened linewidths. For our analysis of 
these data we assigned the H$_2$-He-broadened linewidth for all lines to be 
0.101 cm$^{-1}$, which is an average of the H$_2$-broadened linewidths of lines 
in this band reported by Keffer et al. (1985, 1986). However, a key drawback of 
these data is that they are derived from room temperature observations only and thus do not 
include an estimate of the lower state energy of the lines. Hence, it 
is again difficult to use these data to model the infrared spectrum of cold planets 
such as Jupiter or hot exoplanets since we cannot compute how the strengths vary 
with temperature. 

\begin{figure}
\epsscale{0.8}
\plotone{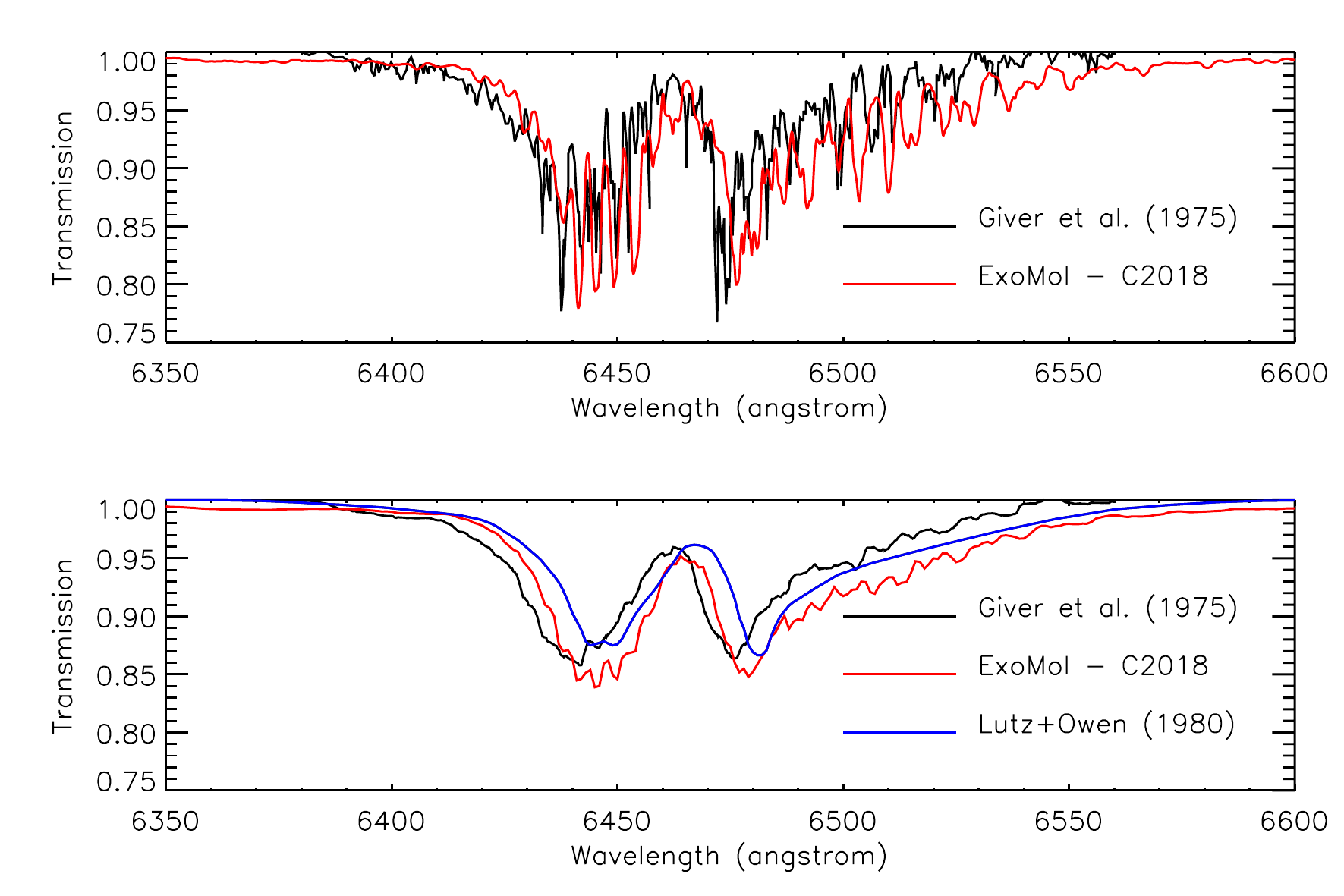}
\caption{Top: Transmission spectrum of the self-broadened path of ammonia shown in Fig. 3 of \cite{giver75}. In this calculation the temperature is 294 K, the path length is 36 m and the pressure is 1 atm. The originally measured `low-dispersion' spectrum of \cite{giver75} (digitised by the authors) is shown in black. The red line in this panel shows the spectrum calculated using the new ExoMol--C2018 line data of \cite{jtNH3PES}. The resolving power of this spectrum was not stated, but we found that if we set it to R = 3000 (i.e. d$\lambda = 2.15$ \AA) we achieved reasonably good correspondence with the observed spectrum, although the line centres do not always align. Bottom: Transmission spectrum of the same self-broadened path of ammonia  calculated using the absorption coefficients of Lutz and Owen (1980) (blue line). The authors claim these coefficients have a resolution of 2 \AA\ (or 0.0002 $\mu$m), but comparing with spectra calculated using the new ExoMol--C2018 line data of \cite{jtNH3PES}, we find that a smoothing the ExoMol data with a square function of full-width-half maxium (FWHM) of 10 \AA\ (red line) gives a spectrum that compares much more favourably with that calculated with the coefficients of \cite{lutzowen80}. In this bottom plot we also show the  `low-dispersion' spectrum of \cite{giver75} smoothed to the same resolution of 10 \AA\ (black) for reference. \label{fig6475}}
\end{figure}

Fig. \ref{fig6475} shows a `low dispersion' spectrum of a laboratory path of 
ammonia measured by \cite{giver75}, where we hand-digitised the spectrum shown 
in their Fig.3 and which we previously showed in Fig. 2 of \cite{irwin18}. The 
path was 36 m long and contained 1 atm pressure of pure ammonia at a temperature 
of 294 K. In Fig. \ref{fig6475} we compare the measured spectrum 
with that calculated using the new ExoMol--C2018  line data \citep{jtNH3PES}, where we found that smoothing the line-by-line calculated spectrum to a spectral resolution of R = 3000 (i.e. 2.15 \AA)
provided the best correspondence to the measured Giver et al. spectrum. In Fig. \ref{fig6475} we also show the corresponding spectrum calculated with the cross-section data of \cite{lutzowen80}. The resolution of these data is quoted as 2 \AA\ by \cite{lutzowen80}, but when we smoothed our line-by-line calculated spectrum to this resolution we found poor correspondence. Instead, in Fig. \ref{fig6475}, we smoothed the C2018-calculated line-by-line spectrum to a resolution of 10 \AA\ , which we found provided a reasonable match to the resolution of the spectrum 
calculated with the Lutz\&Owen data, and conclude that the true 
resolution of the Lutz\&Owen data is actually 10 \AA, not 2 \AA.

\begin{figure}
\epsscale{0.6}
\plotone{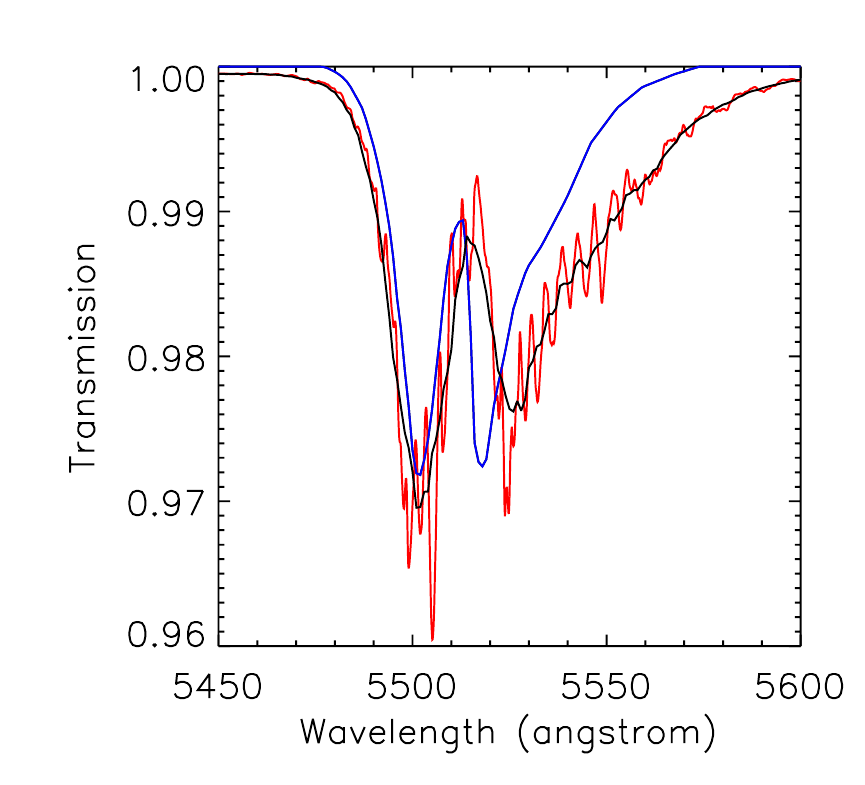}
\caption{Calculated transmission for the same path computed in Fig. \ref{fig6475}, but centred on the absorption band at 5520 \AA. The blue line is the spectrum calculated using the \cite{lutzowen80} absorption coefficients. The red line is the spectrum calculated with the ExoMol--C2018 line data \citep{jtNH3PES}, smoothed to a spectral resolving power of 3000 (i.e. d$\lambda$ = 1.8\AA), while the black line is the same spectrum smoothed to a resolution of 10 \AA, which is the believed resolution of the Lutz\&Owen data. \label{fig5520}}
\end{figure}

It is clear from Fig. \ref{fig6475} that the ExoMol--C2018 ammonia line data \citep{jtNH3PES} 
provide a reasonable correspondence, given the uncertainties in line positions of the ExoMol line data,
with the self-broadened laboratory 
measurements of \cite{giver75} and \cite{lutzowen80} for the 6475 \AA\ band.  
We also looked to see how well these data matched the data of \cite{lutzowen80} for 
the 5520 \AA\ band (this band was not measured by \cite{giver75}). Figure 
\ref{fig5520} shows the spectrum of the 5520 \AA\ band calculated for the same 
path length and conditions as Fig. \ref{fig6475} using the \cite{lutzowen80} 
data. Overplotted in Fig. \ref{fig5520} is the spectrum calculated with the 
C2018 line data, smoothed to a resolution of 1.8 \AA\ (i.e. R = 3000) and 10 
\AA, respectively. We find that the lower resolution calculated spectrum 
provides a reasonable match to the Lutz\&Owen spectrum, but that there are 
differences in the calculated position of the longer-wavelength absorption peak. It 
is not clear whether this discrepancy arises from: 1) errors in digitising the 
Lutz\&Owen data; 2) errors in the wavelength scale of the published Lutz\&Owen 
data; or 3) errors in the line positions in the C2018 line data \citep{jtNH3PES}. In particular,
the potential energy surface used to construct the line list was
only tuned to energy levels below 10,000 cm$^{-1}$ and can therefore be
expected to become increasingly inaccurate for wavelengths shorter than 1 $\mu$m. 
However, we also note that there appears to be a discrepancy between Figs. 1 and 4 of \cite{lutzowen80}.
In Fig. 1 of \cite{lutzowen80}, the shorter wavelength absorption peak is clearly at a wavelength less than 5500 \AA\, and the second peak at a wavelength least 25 \AA\ longer.  
However, in their Fig. 4, it appears that there is only a 56 cm$^{-1}$ difference between the peaks, which equates to a 17 \AA\ difference. 
Hence, it may be that there are indeed errors in the wavelength scale of the published Lutz\&Owen data.

\begin{figure}
\epsscale{1.0}
\plotone{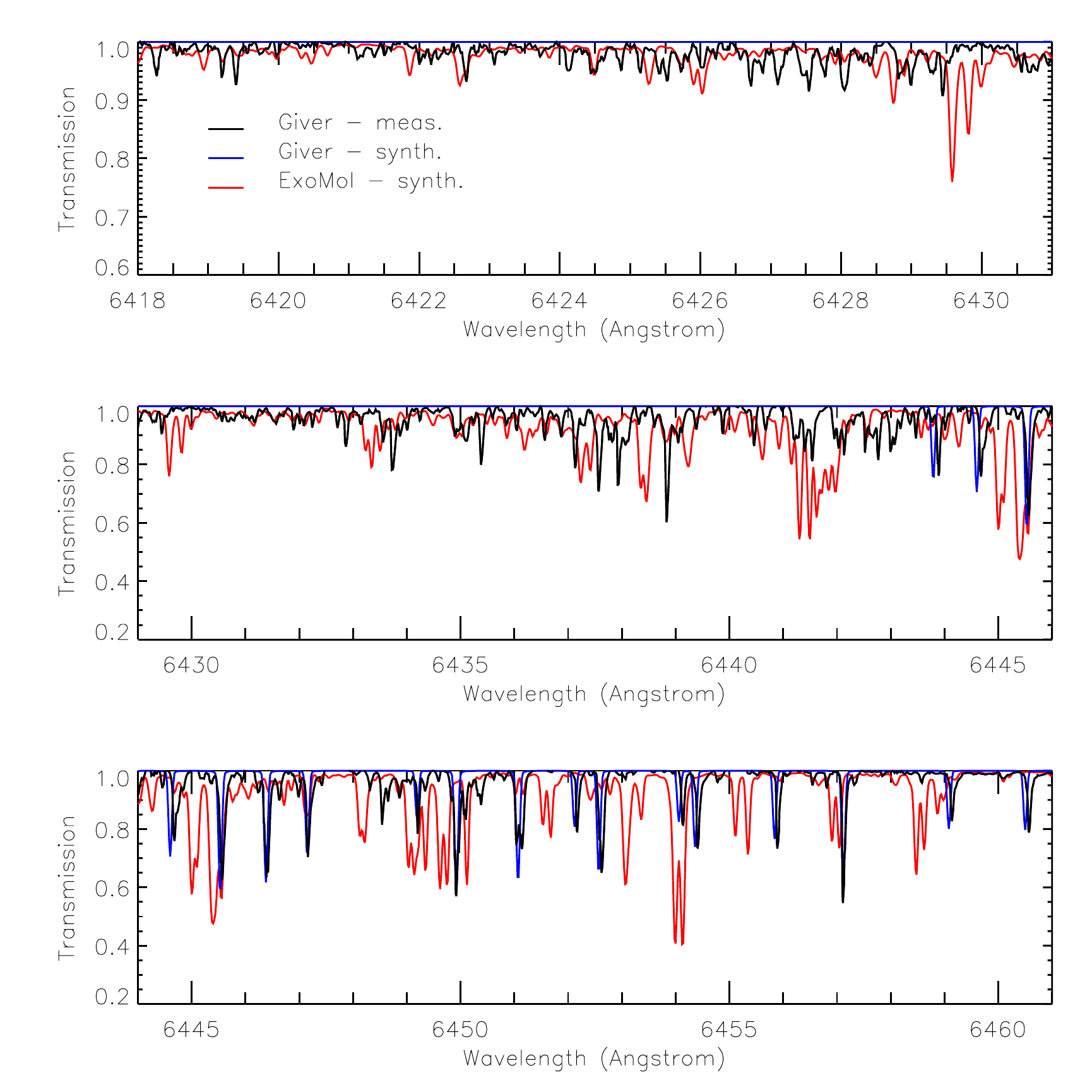}
\caption{Comparison of high resolution spectra of \cite{giver75} (their Fig. 1) with high spectral resolution line-by-line calculations using the line data of \cite{giver75} and ExoMol--C2018 \citep{jtNH3PES} for the conditions specified in Fig. 1. of \cite{giver75}, i.e. length = 400m, p = 0.061 atm, T = 294 K. Here the black lines are the observed spectra in the different spectral ranges (digitised by the authors), the blue lines are the spectra calculated from the Giver et al. line data, while the red lines are the spectra calculated from the ExoMol--C2018 line data. Here the three spectra cover the spectral range 6418--6460 \AA. \label{sym1}}
\end{figure}

\begin{figure}
\epsscale{1.0}
\plotone{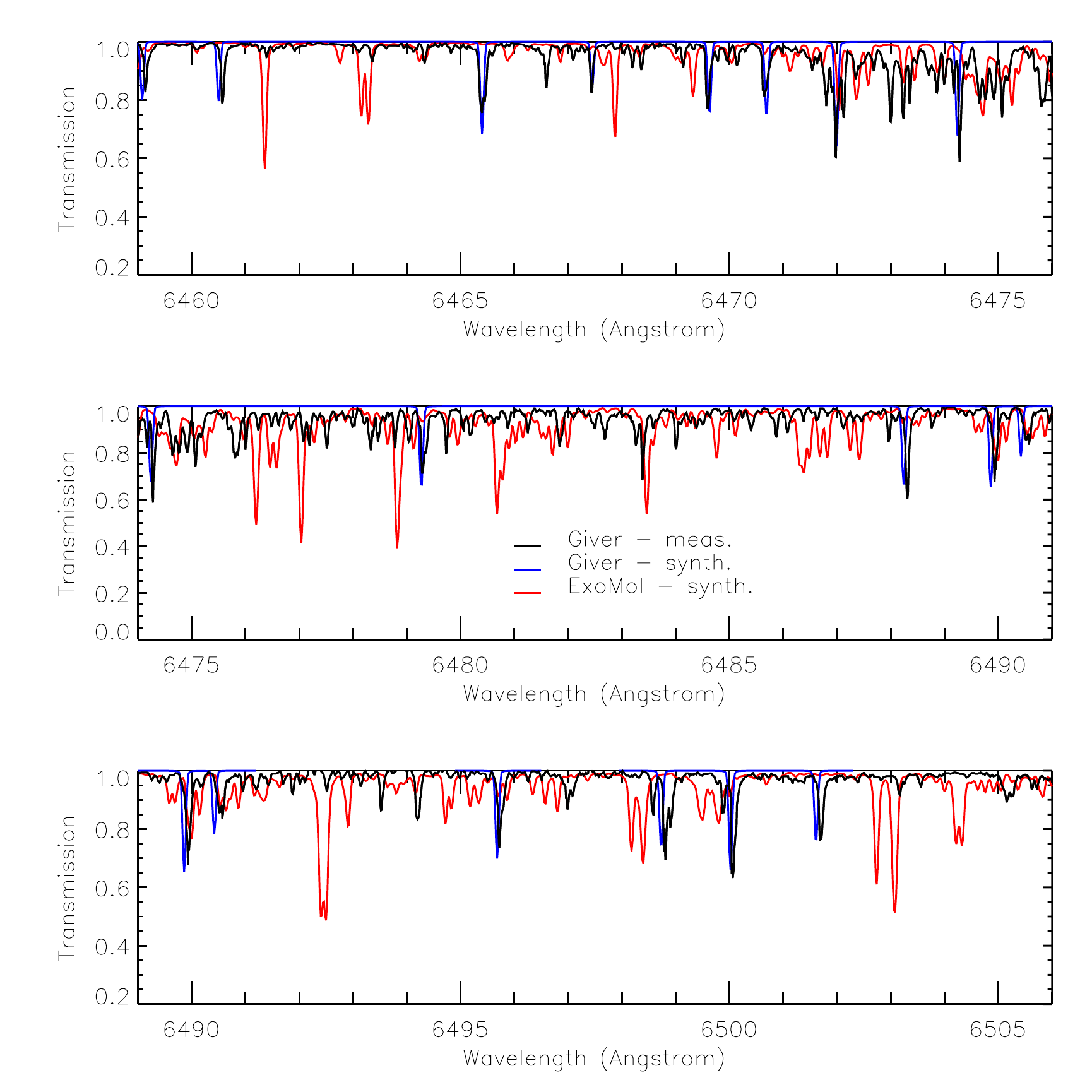}
\caption{As Fig. \ref{sym1}, but here covering the spectral range  6460--6505 \AA. \label{sym2}}
\end{figure}

\begin{figure}
\epsscale{1.0}
\plotone{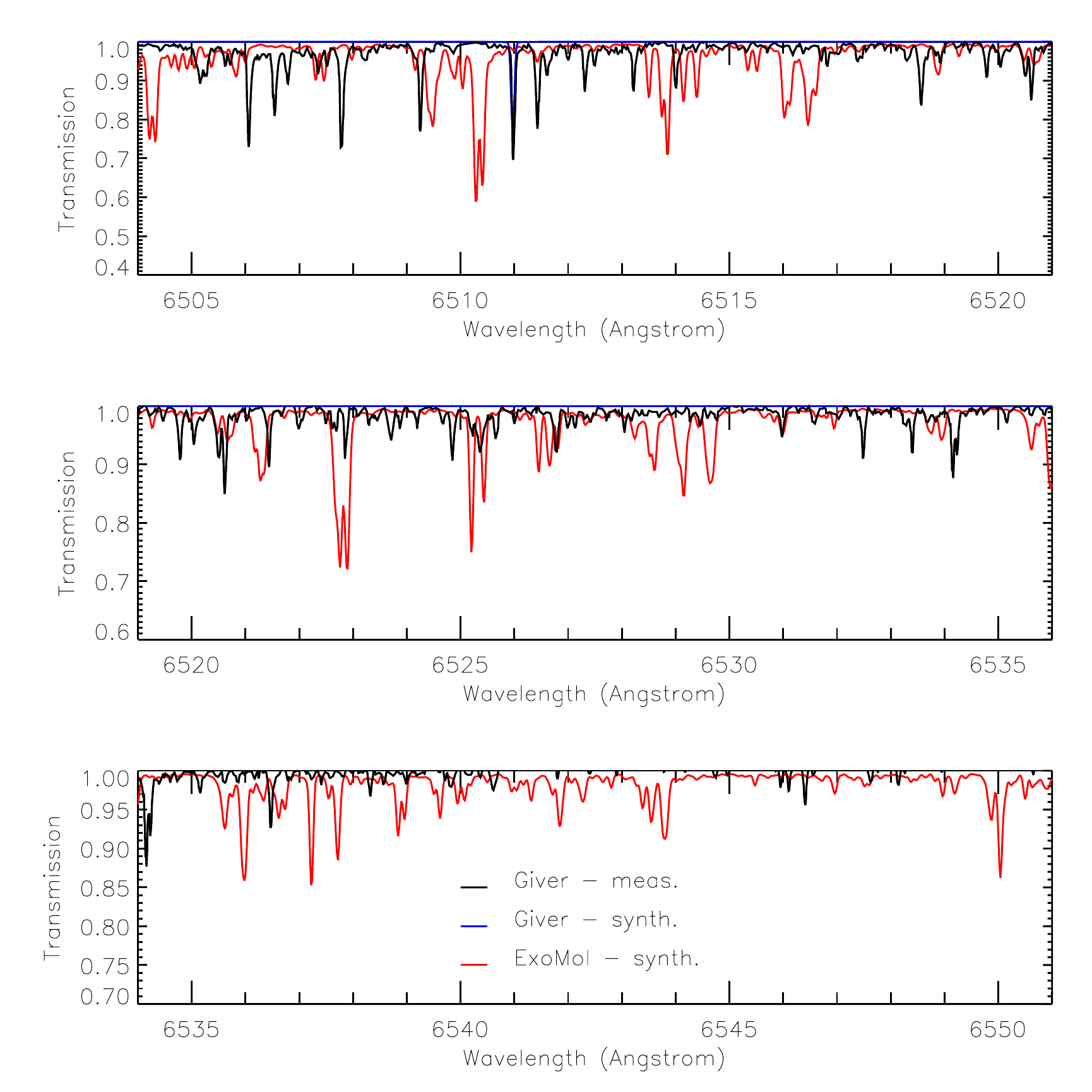}
\caption{As Fig. \ref{sym1}, but here covering the spectral range  6505--6550 \AA. \label{sym3}}
\end{figure}

\begin{figure}
\epsscale{0.9}
\plotone{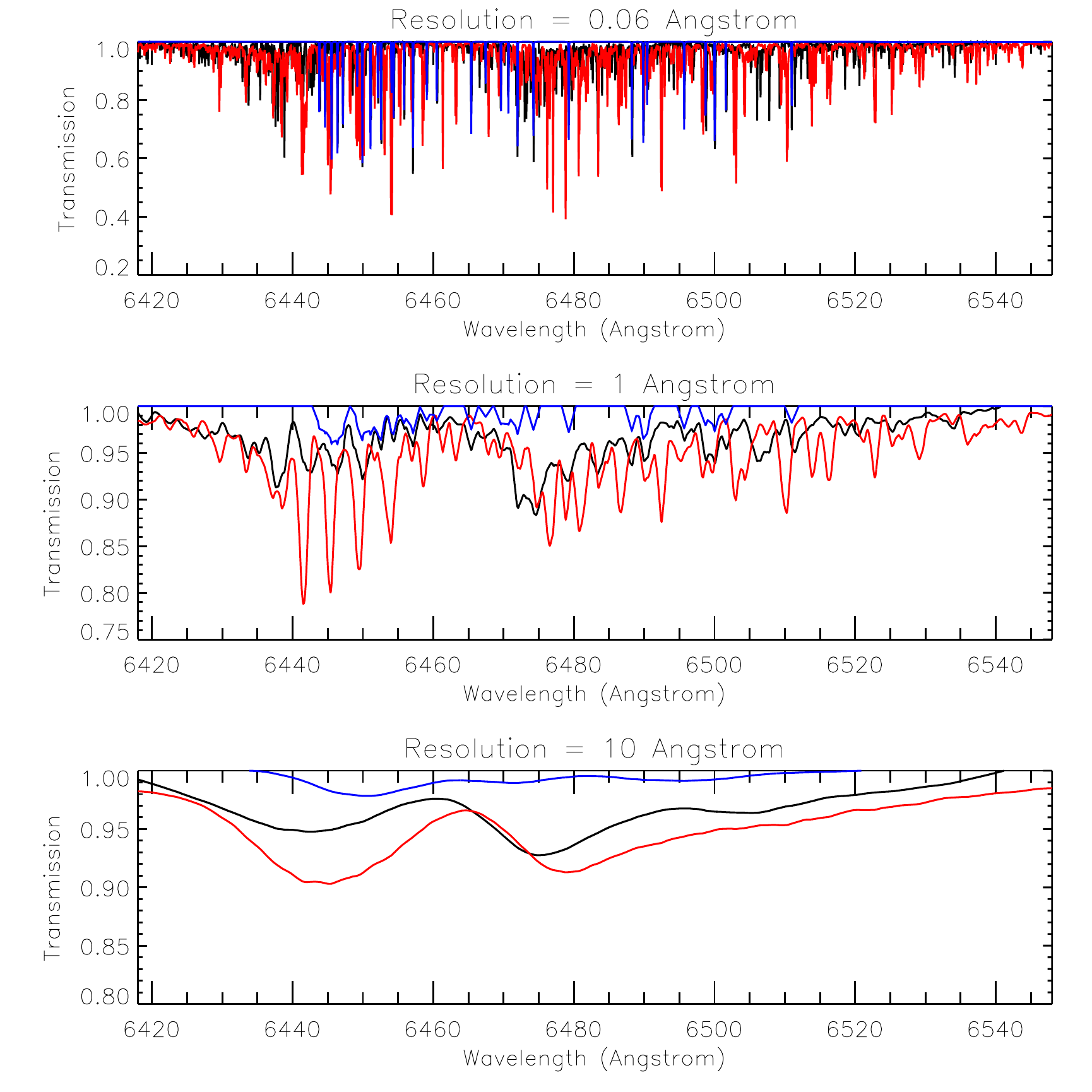}
\caption{Comparison of high resolution spectra of \cite{giver75} covering the whole 6418 -- 6548 \AA\ range, with high spectral resolution line-by-line calculations using the line data of \cite{giver75} and ExoMol--C2018 \citep{jtNH3PES} for the conditions specified in Fig. 1. of \cite{giver75}, i.e. length = 400m, p = 0.061 atm, T = 294 K, and different modelled spectral resolutions (assumed triangular line shape) with FWHM = 0.06, 1.0 and 10.0 \AA\, respectively. Here the black lines are the observed spectra in the different spectral ranges (digitised by the authors), the blue lines are the spectra calculated from the Giver et al. line data, while the red lines are the spectra calculated from the ExoMol--C2018 line data.  \label{symsum}}
\end{figure}

To compare the line data of the ExoMol--C2018  line database with the measurements 
in more detail, we hand-digitised the published high-resolution (quoted spectral resolving power R $\sim$ 170,000) 
laboratory spectra of Giver et al. for the 6475 \AA\ band \citep[shown as Fig.1 of][]{giver75} for a low pressure path of length 400m, pressure 0.061 atm and temperature 294K. \cite{giver75} showed these spectra spread over three pages and we repeat this format here, comparing our digitised version with line-by-line-calculated spectra in Figs \ref{sym1} -- \ref{sym3}. 
We find that spectra calculated\footnote{N.B. We found that our line-by-line spectra had to be smoothed to a resolution of 0.06 \AA\ to provide best correspondence with the observed spectra, which indicates that the actual spectral resolution of the high resolution observations was $\sim$108,000, rather than $\sim$170,000 as quoted by \cite{giver75}} with the line data of \cite{giver75} have absorption lines that match the position and strength of many of the stronger measured lines, for example at 6449.95 \AA\ and 6457.05 \AA, but that many are not represented; this is entirely as expected since these line data were only published for the thirty-three strongest lines. Spectra calculated from the ExoMol--C2018 line data match better the overall distribution of different 
line strengths, (i.e. that in broad wavelength intervals the ExoMol lines are weak where the measured lines are weak, and the ExoMol lines are strong where the measured lines are strong) but the line centres do not match the individual observed line positions very well at all. The generally good agreement of the ExoMol lines with broad distribution of observed absorption lines can be seen more clearly in Fig. \ref{symsum}, where the measured absorption spectrum is compared with the calculated spectra over the entire 6418 -- 6548 \AA\ wavelength range at three different spectral resolutions: 0.06 \AA (i.e. original resolution), 1.0 \AA\ and 10.0 \AA\ (all assuming a triangular line shapes).  
We can see that the spectra calculated with the ExoMol line data reproduce the overall shape of the absorption band well at lower resolution, but predict slightly more absorption and also predict that the wavelengths of peak absorption occur at slightly longer wavelengths than those observed (5 -- 10 \AA). The same comparisons can be drawn when comparing the spectra observed/calculated for the higher pressure path shown in Fig. \ref{fig6475}. At high resolution, the poor correspondence between the observed and predicted line centres, but better correspondence between the observed and predicted line strengths is not surprising since at shorter wavelengths we expect these computed data to be more accurate in predicting the strengths of different absorption lines than their line 
positions. Owing to the lack of empirically-derived energies below 1 $\mu$m at the time of producing the PES used by \cite{jtNH3PES}, line positions below 1 $\mu$m may be several, or even tens of wavenumbers in error depending on the upper state vibrational mode and $J$. These inaccuracies are discussed by \cite{jt715} in their analysis of the red and green optical spectrum of ammonia. Hence, we conclude that the ExoMol--C2018 line data provide a very useful resource for modelling low- to medium-resolution observed spectra, but should be used with caution when analysing high-resolution observations.  

\section{Comparison with Bowles et al. (2008) absorption data}

\begin{figure}
\epsscale{0.9}
\plotone{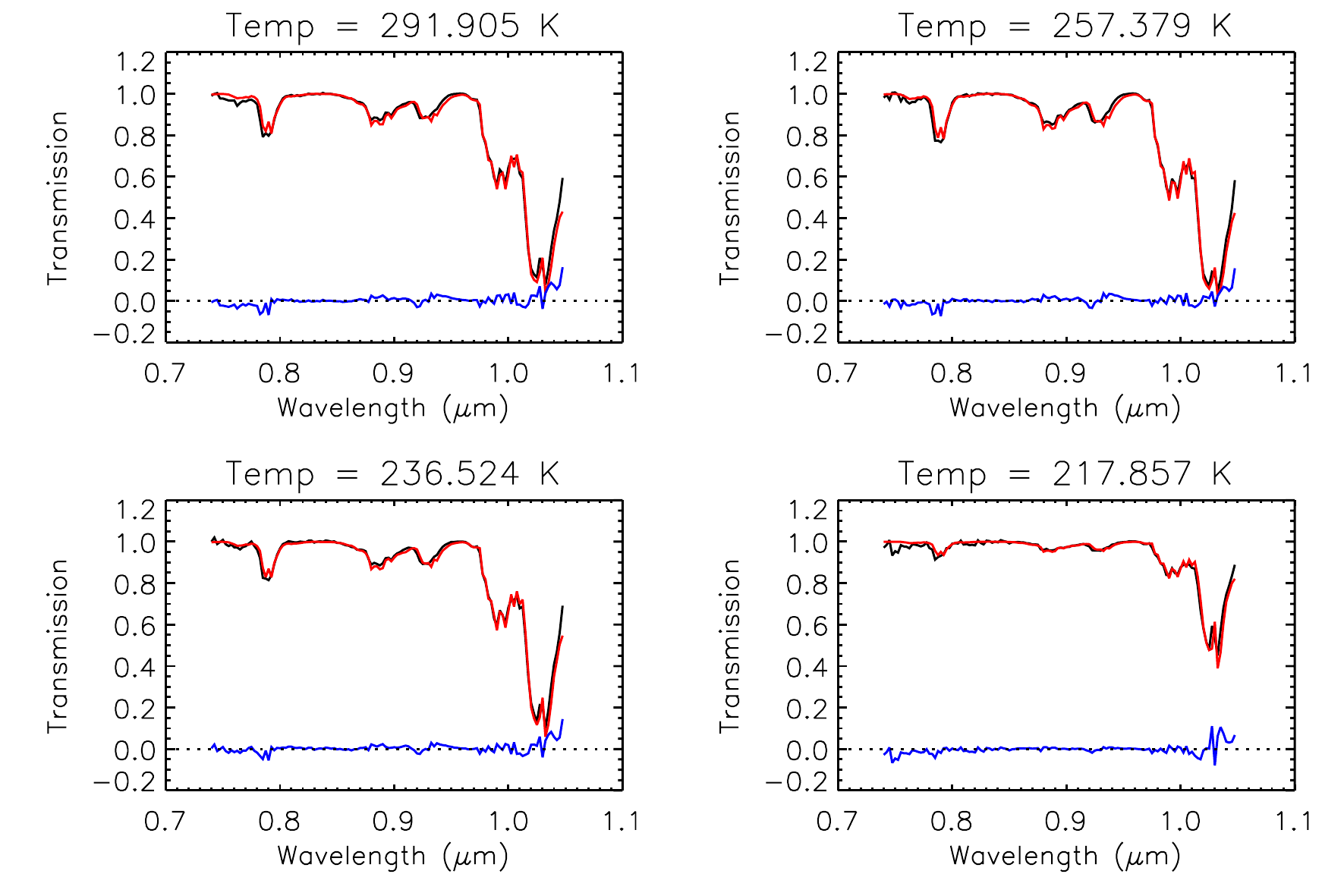}
\caption{Laboratory-measured transmission spectra (black) for the longest paths observed by \cite{bowles08} at a range of temperatures and shown in Fig. 4 of \cite{irwin18}, compared with spectra calculated with the ExoMol--C2018 line data \citep[red,][]{jtNH3PES} at the same resolution (0.0025 $\mu$m), and difference (blue). We can see that there is excellent correspondence and also that the spectra calculated with the C2018 line data are much less noisy than the observed spectra, particularly at the shortest wavelengths  \label{bowlestran}}
\end{figure}

\cite{bowles08} report Goody-Lorentz band models fitted to the measured 
absorption spectra of laboratory paths of ammonia with path lengths of 2.164 and 
10.164 m, pressures ranging from 0.075 to 1.02 bar and temperatures varying 
between 216 and 292 K. The measured transmission spectra cover the spectral 
range 0.74 - 1.0475 $\mu$m and have a spectral resolution and step of 0.0025 
$\mu$m. \cite{irwin18} found that the band data of \cite{bowles08}  become 
increasingly noisy at shorter wavelengths. The shortest wavelength absorptions 
were only visible at the longest path lengths, highest pressures and highest 
temperatures in the laboratory measurements and these features were essentially 
lost in the noise at lower temperatures. This is because at lower temperatures, 
since the pressure cannot be allowed to exceed the saturated vapour pressure of 
ammonia (to avoid condensation on the surfaces of the experiment) this 
necessarily limits the path amounts that can be attained. Figure 
\ref{bowlestran} shows the laboratory spectra measured by \cite{bowles08} at the 
longest available path lengths for four temperatures in the range 215 -- 295 K 
(shown as Fig. 4 in \cite{irwin18}), together with simulated spectra for the 
same paths calculated with the ExoMol--C2018 line data (line-by-line) for self-broadened conditions and then 
smoothed to the resolution of the \cite{bowles08} data, i.e. 0.0025 $\mu$m. As 
can be seen there is an excellent correspondence between the measured spectra and the 
ExoMol--C2018 calculations, with mean transmission differences of 0.015, increasing to 0.15 at the very longest wavelength. However, the ExoMol--C2018 \citep{jtNH3PES} line data has a much better, smoother variation at the shorter wavelengths, where the Bowles data was 
found by \cite{irwin18} to become unreliable.  

\section{Analysis of MUSE observations}
\begin{figure}
\epsscale{0.7}
\plotone{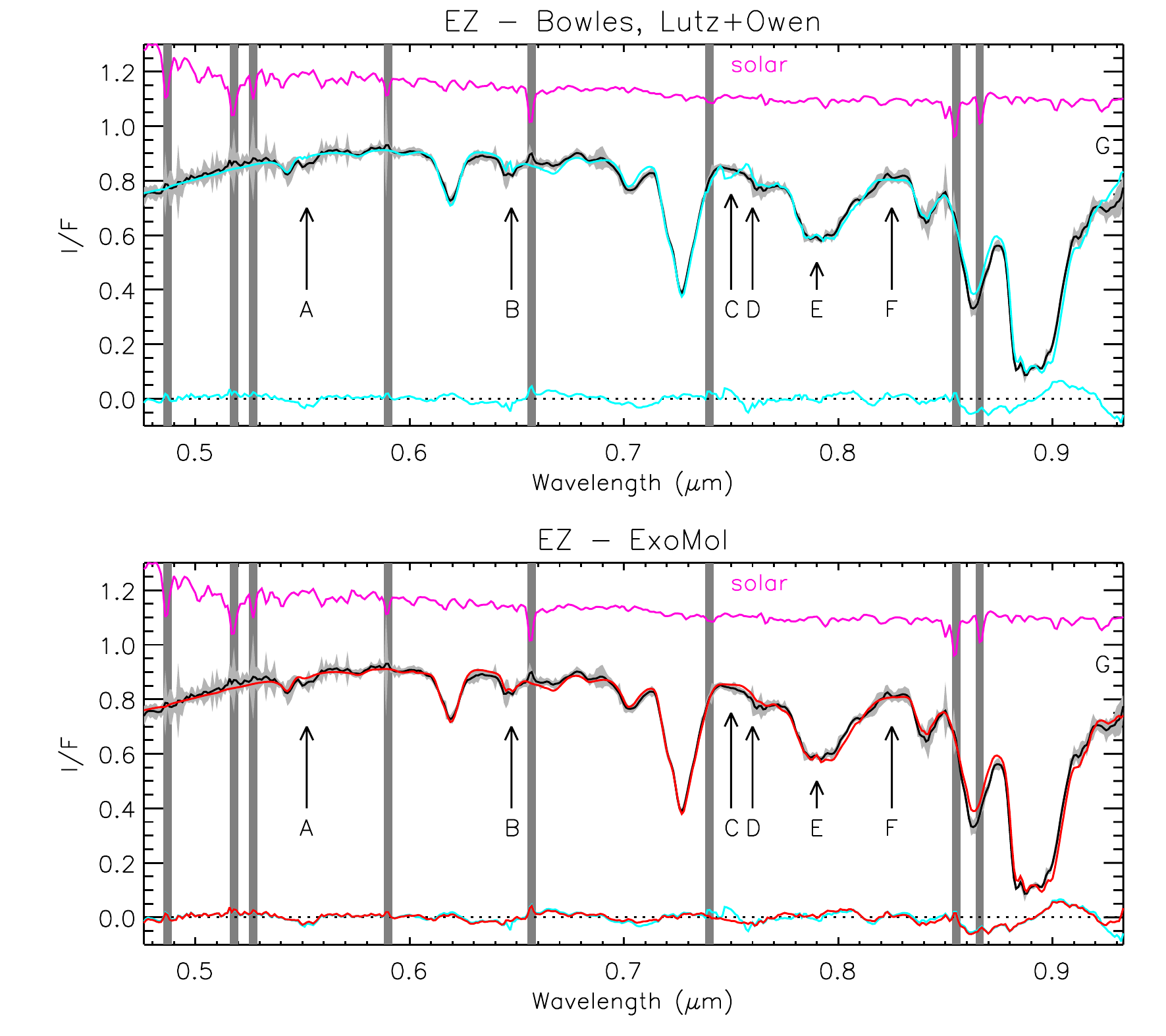}
\caption{Observed MUSE spectrum taken from a single pixel in the Equatorial Zone (EZ) with estimated error (including forward model errors) shown in grey, observed on 9th April 2018 at 06:04:08UT (similar to Fig.9 of \cite{irwin18}, which shows a fit to data from 28th May 2017). In the top panel, the cyan line is the fit of our NEMESIS retrieval model using the ammonia absorption data of \cite{bowles08} and \cite{lutzowen80}. In the bottom panel, the red line is the fit of our NEMESIS retrieval model using the new ExoMol--C2018 ammonia line data of \cite{jtNH3PES}. The plots also show the differences, with the bottom plot also showing the difference (cyan) when using the previous ammonia absorption data of \cite{bowles08} and \cite{lutzowen80}. The labelled features `A' to `G' mark the main observable ammonia absorption features in the ammonia absorption data of \cite{bowles08} and \cite{lutzowen80}. Also shown in magenta is the assumed solar spectrum of \cite{chance10}, divided by a Planck function of temperature 5778 K and scaled to give a value of $\sim 1.0$, showing the various Fraunhofer lines in the spectrum, which are highlighted by the vertical grey bars. \label{jup_ez}}
\end{figure}

Figure \ref{jup_ez} shows a typical spectrum of Jupiter extracted from our MUSE observations \citep{braude19} from 9th April 2018 from a single pixel in the Equatorial Zone near the sub-Earth point and similar to the spectrum observed on 28th May 2017 
shown as Fig. 9 of \cite{irwin18}. As before the MUSE spectra were smoothed with a triangular line shape of FWHM=0.002 $\mu$m, and sampled with a step of 0.001 $\mu$m, giving a 
spectral resolution of $R \sim 250$. Similarly, as in our previous analysis, these spectra 
were modelled with our NEMESIS \citep{irwin08} radiative transfer and retrieval 
tool, using the method of correlated-k (e.g. \cite{lacisoinas91}) in its 
radiative transfer scheme, and modelling multiple scattering using the 
matrix-operator method of \cite{plass73}. The absorption of ammonia gas was 
modelled using either the band data of \cite{bowles08}, converted to k-tables as 
described by \cite{irwin18}, together with the  \cite{lutzowen80} cross-sections 
for the  0.55 and 0.65 $\mu$m ammonia absorption bands (with no temperature 
dependance assumed), or using the new k-tables generated from the ExoMol--C2018 line list \citep{jtNH3PES}, described earlier. 
Methane absorption was modelled with a k-table generated from the band data of 
\cite{kark10}, while small absorption features from H$_2$ lines were modelled 
with a k-table generated from line data from the HITRAN 2012 line database 
\citep{rothman13}. The collision-induced absorption of H$_2$--H$_2$ and H$_2$--He at these 
wavelengths was modelled after \cite{borysow89a,borysow89b,borysow00}, while at shorter 
wavelengths, the Rayleigh-scattering opacity was modelled via standard theory \citep[e.g.][]{goody89,hansen74}, 
accounting for the contributions of H$_2$, He, CH$_4$ and NH$_3$. For this 
reflectance calculation, the observed radiances were divided by a reference 
solar spectrum, using Jupiter's distance from the sun of 5.42~AU on this date. 
For this, the solar spectrum of \cite{chance10} was used, which was first 
smoothed with a triangular line shape of FWHM = 0.002 $\mu$m to make it 
compatible with our smoothed spectra. 

\begin{figure}
\epsscale{0.7}
\plotone{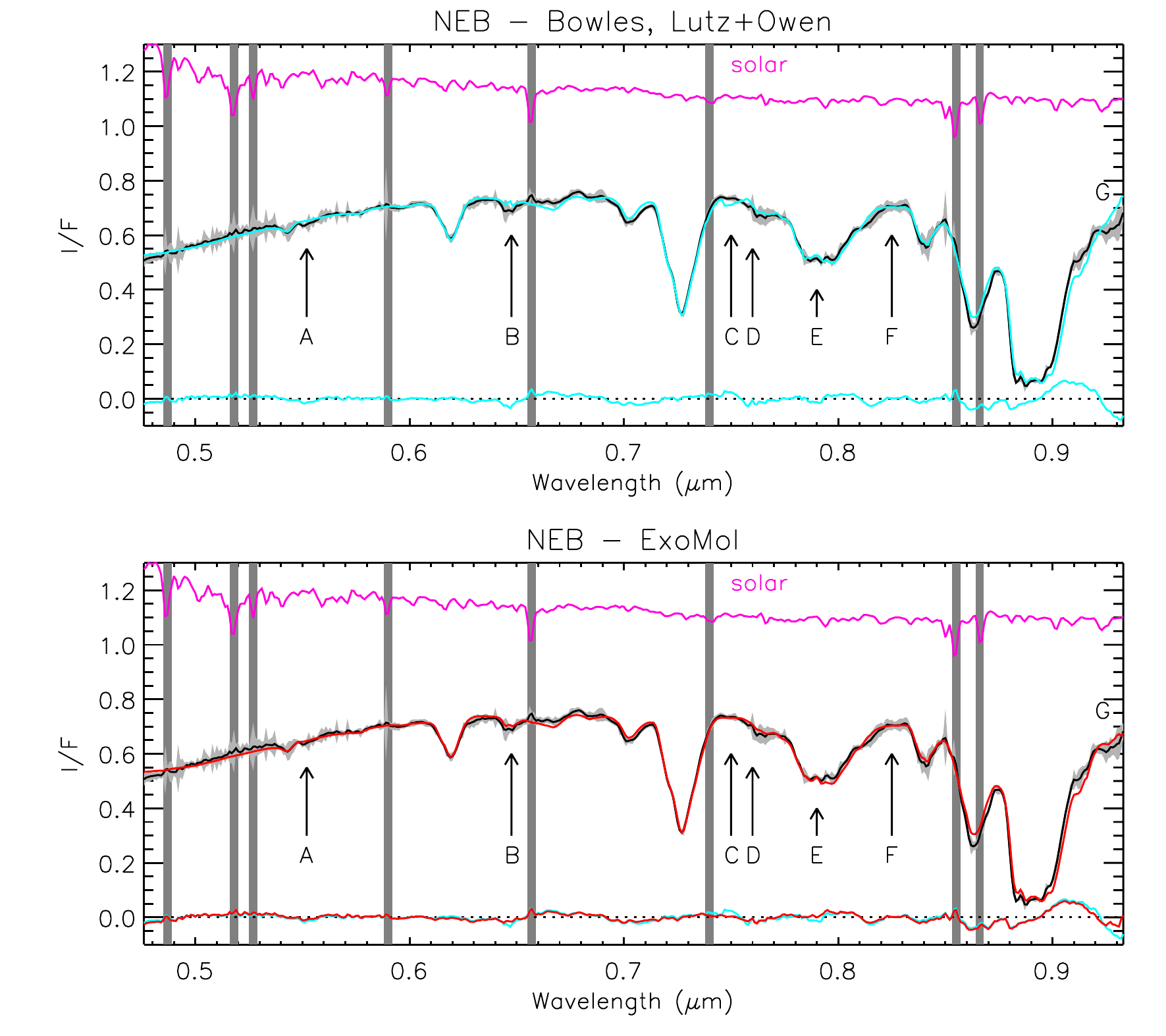}
\caption{As Fig. \ref{jup_ez}, but comparing our calculations for different ammonia absorption sources with the observed MUSE spectrum in a single pixel in the North Equatorial Belt (NEB).  \label{jup_neb}}
\end{figure}

Figure \ref{jup_ez} shows our best fit to the Equatorial Zone spectrum using the 
combined Bowles et al. and Lutz\&Owen data, and then the fit resulting from using the NH$_3$ k-table generated from the 
ExoMol--C2018 line list \citep{jtNH3PES}. Figure \ref{jup_neb} shows the same 
comparison for a spectrum measured in the North Equatorial Belt (corresponding 
to Fig. 10 of \cite{irwin18}). 

As can be seen we are able to fit the VLT/MUSE observations of the EZ and NEB  significantly better with the ExoMol--C2018 NH$_3$ absorption data than we can using our previous best combination of the data of  \cite{bowles08} and \cite{lutzowen80}. The $\chi^2/n$ of the fit was reduced from 1.93 to 1.60 for the EZ, and from 2.06 to 1.76 for the NEB. Our previous list of visible NH$_3$ `features' are marked in Figs. \ref{jup_ez} and \ref{jup_neb}  as: A -- 0.552 $\mu$m, B -- 0.648  $\mu$m, C -- 0.75 $\mu$m, D -- 0.765 $\mu$m, E 
-- 0.79 $\mu$m, F -- 0.825 $\mu$m and G -- 0.93 $\mu$m. It can be seen that the feature `C' now appears to be an artefact of the data of \cite{bowles08}, but the others seem real and are clearly very well fit simultaneously using the ExoMol--C2018 line data. We should also that feature `D' at 0.765 $\mu$m coincides almost exactly with an O$_2$-absorption band in the telluric spectrum. This appears to have been incompletely corrected for in the spectrum from 2017 shown by \cite{irwin18}, which made the discrepancy at this wavelength look worse. For the spectrum presented here (from 2018) where a more reliable telluric correction was achieved, feature `D' is much less pronounced and seems equally well fit by the ExoMol--C2018 and Bowles et al. data. However, in summary, it can be seen that over the whole MUSE spectral range, the new ExoMol--C2018 line data for NH$_3$ \citep{jtNH3PES} greatly improve our ability to model the near-infrared spectrum of Jupiter.

\section{Conclusion}
The ExoMol--C2018 NH$_3$ line list of \cite{jtNH3PES} clearly represents a significant step forward in allowing the accurate modelling of ammonia absorption in the visible/near-IR spectra of giant planets. However, while the ExoMol--C2018 data is shown to reproduce low to medium
resolution features well ($R \sim 250$ and below), it is clear that the accuracy to which the wavelengths  of individual lines
are predicted needs to be improved for high resolution studies. It is possible
to do this {\it post hoc} (e.g. \cite{jt570}), provided either high-resolution
measured frequencies or empirical energy levels are obtained. 
High resolution laboratory spectroscopy studies of ammonia have long been
available, e.g. \cite{88LeCo}; however, coverage remains fragmentary and 
analysis of these spectra difficult, as noted by \cite{jt715}. New laboratory studies
in this region would be welcome and would facilitate further improvement of
the available line lists.
 



\section{Acknowledgements}

The VLT/MUSE observations were performed at the European Southern Observatory (ESO), proposals: 095.C-0149, 096.C-0173, 098.C-0035, 099.C-0192 and 101.C-0097. We thank Larry Sromovsky for kindly providing the code we used to generate our Rayleigh-scattering opacities, Renyu Hu, for providing an electronic version of the line data of \citep{giver75} and advising on the use of the linewidths of \cite{keffer85} and \cite{keffer86}, and Sergi Hildebrandt and Margaret Turnbull (WFIRST Science Investigation Team PI) for coordinating the WFIRST Exoplanet Data Challenge, through which these discrepancies in ammonia absorption coefficients came to light. We would also like to thank our anonymous reviewers for their very detailed and helpful reviews.
The work at University College London was supported by the UK Engineering and Physical Sciences  Research Council (EPSRC)
grant EP/M506448/1 and Servomex Ltd,
the use of the UCL Legion High Performance Computing Facility (Legion@UCL), 
along with the STFC DiRAC HPC Facility supported by BIS National E-infrastructure capital grant ST/J005673/1 
and STFC grants ST/H008586/1, ST/K00333X/1.



{\it Facilities:}   \facility{VLT (MUSE)}.

\clearpage



\end{document}